\documentclass[a4paper]{article}

\usepackage[utf8]{inputenc}

\usepackage{multicol}
\usepackage{amsthm}
\usepackage{bm}

\usepackage[dvipdfmx]{graphicx}

\usepackage{bmpsize}
\usepackage{amssymb}
\usepackage{amsmath}
\usepackage{color}
\usepackage{amsthm}

\newcommand{\BA}{\begin{eqnarray}}
\newcommand{\EA}{\end{eqnarray}}

\definecolor{dgreen}{rgb}{0.0, 0.5, 0.0}

\begin{document}

\fontsize{14pt}{16.5pt}\selectfont

\begin{center}
\bf{Relation of stability and bifurcation properties\\ between continuous and ultradiscrete dynamical systems\\  via discretization with positivity: one dimensional cases}
\end{center}
\fontsize{12pt}{11pt}\selectfont
\begin{center}
Shousuke Ohmori$^{*)}$ and Yoshihiro Yamazaki\\ 
\end{center}

\noindent
\it{Department of Physics, Waseda University, Shinjuku, Tokyo 169-8555, Japan}\\

\noindent
*corresponding author: 42261timemachine@ruri.waseda.jp\\
~~\\
\rm
\fontsize{11pt}{14pt}\selectfont\noindent

\baselineskip 30pt

{\bf Abstract}\\
%
Stability and bifurcation properties of one-dimensional discrete dynamical systems with positivity, which are derived from continuous ones by tropical discretization, are studied.
%
%
%
The discretized time interval is introduced as a  bifurcation parameter in the discrete dynamical systems, and emergence condition of an additional bifurcation, flip bifurcation, is identified.
%
Correspondence between the discrete dynamical systems with positivity and the ultradiscrete ones derived from them is discussed. 
It is found that the derived ultradiscrete max-plus dynamical systems can retain the bifurcations of the original continuous ones via tropical discretization and ultradiscretization.




%
%
\section{Introduction}

Tropical discretization and ultradiscretization are one of approaches to the Wolfram's 9th problem\cite{Wolfram1985}, which refers to correspondence between continuous dynamical systems and discrete ones such as cellular automata.
Recently, this approach has been applied to various dynamical systems such as SIR model\cite{Willox2003}, a model for an inflammatory response\cite{Carstea2006, Willox2007}, 
Allen-Cahn equation\cite{Murata2013}, 
Gray-Scott model\cite{Matsuya2015}, 
a model for biological rhythms\cite{Gibo2015}, 
a reaction-diffusion model\cite{Ohmori2016}, 
normal forms in one dimensional dynamical systems\cite{Ohmori2020}, 
Sel'kov model\cite{Yamazaki2021,Ohmori2021}, and van der Pol equation\cite{Isojima2022}.

%
%
%
%
%
%
%
\color{black}

Tropical discretization is a discretizing procedure converting a differential equation into a difference equation with only positive variables\cite{Murata2013}.  
Let us consider a differential equation of $x$,  
\begin{eqnarray}
	\frac{dx}{dt}& = F(x)=f(x)-g(x),
	\label{eqn:1-1} 
\end{eqnarray} 
where we assume that $x(t)> 0$ and that $F(x)$ can be divided into the two positive smooth functions $f$ and $g$.
Then, as the tropical discretization of eq. (\ref{eqn:1-1}), 
the following discretized form is adopted:
\begin{eqnarray}
	x _{n+1}=x _n \frac{x _n+\tau f(x _n)}{x _n+\tau g(x _n)} \equiv F_\tau (x_n), 
	\label{eqn:1-2}
\end{eqnarray}
where $x_n = x(n \tau)$; 
$\tau ( > 0 ) $ and $n$ show the discretized time interval and 
the number of iteration steps, respectively.
It is noted that eq. (\ref{eqn:1-1}) is reproduced  
from the following equation that is identical to eq. (\ref{eqn:1-2}),
	\begin{eqnarray}
		\displaystyle\frac{x _{n+1}-x _n}{\tau }= x _n \frac{f(x _n)-g(x _n)}{x _n+\tau g(x _n)}
		\label{eqn:1-3}
	\end{eqnarray}
by taking $\tau \rightarrow0$. 
%
%
The tropical discretization is also known 
as a non-standard finite difference scheme 
with positively-preserving system of ordinary differential equations\cite{Mickens1994,Alexander2005}.

Ultradiscretization is a limiting procedure transforming a difference equation
into another type of difference equation with max-plus algebra\cite{Tokihiro1996}.
First, for some positive variables $a, b, \ldots$, they are transformed into  $A, B, \ldots$ by $a=e^{A/\varepsilon}, b=e^{B/\varepsilon}, \ldots$, where $\varepsilon$ is a positive parameter.
Next, after this transformation, the following ultradiscrete limit is executed:
\begin{eqnarray}
	\displaystyle\lim_{\varepsilon  \to +0} \varepsilon  \log(e^{A/\varepsilon }+e^{B/\varepsilon }+\cdot \cdot \cdot )
	=\max(A,B,\cdot \cdot \cdot).		
	\label{eqn:0}
\end{eqnarray}
%
%
%
%
Applying the ultradiscretizing procedure to the tropically discretized equation (\ref{eqn:1-2}) 
after the variable transformations, $x_n=e^{X_n/\varepsilon}, \tau =e^{T/\varepsilon}, f(x_n)=e^{F(X_n)/\varepsilon}, g(x_n)=e^{G(X_n)/\varepsilon}$, 
we obtain the following ultradiscrete equation: 
	\begin{eqnarray}
		X_{n+1}=X_n + \max(X_n,T+F(X_n))-\max(X_n,T+G(X_n)).
		\label{eqn:1-4}
	\end{eqnarray}
%
%
When $T \to \infty$ ($\tau \to \infty$), 
eq. (\ref{eqn:1-4}) becomes
	\begin{eqnarray}
		X_{n+1}=X_n + F(X_n) - G(X_n).
		\label{eqn:1-4a}
	\end{eqnarray}
Therefore, the tropically discretized equation (eq.(\ref{eqn:1-2})) has possibility to formally link the continuous differential equation (eq.(\ref{eqn:1-1})) and the ultradiscrete equation (eq.(\ref{eqn:1-4a})) as the two different limiting cases of $\tau$.

In our previous studies\cite{Ohmori2020,Yamazaki2021,Ohmori2021,Ohmori2022}, 
ultradiscrete bifurcations in eq. (\ref{eqn:1-4a}) have been investigated.
%
%
The important point of our previous results is that 
some ultradiscrete bifurcations coincide with bifurcations of their original differential equations.
For instance, the ultradiscrete equations derived from the one-dimensional normal forms 
of the saddle-node and transcritical bifurcations
possess the ultradiscrete saddle-node and transcritical bifurcations, respectively\cite{Ohmori2020}.
Meanwhile, there are some inconsistent cases.
Actually in the ultradiscrete equation for the supercritical pitchfork bifurcation, 
there exists an additional flip bifurcation which does not exist in the original differential equation.
Such emergence of the additional bifurcation is considered 
to be caused by either the tropical discretization or the ultradiscretization. 
Therefore, it is important to generally identify how the original bifurcations retain and how additional bifurcations emerge; this is the aim of the present manuscript.
%
%
%
%

This paper is organized as follows. 
In the next section, focusing on stability and the local bifurcations, 
we show some results for general relation between continuous differential equations and their tropically discretized ones in one dimension.
Then, we clarify the occurrence condition of the flip bifurcation.
In Sec. 3, based on the general results shown in Sec. 2, 
we review the dynamical properties of the tropically discretized equations we treated previously\cite{Ohmori2020}.
Further in Sec. 4, we discuss correspondence between the discrete dynamical systems with positivity and their ultradiscrete ones.
Conclusion is given in Sec. 5.

\section{General Results}
\label{sec.2}

\subsection{Fixed Point}

%
To begin with, we focus on the relation of fixed points 
between eqs. (\ref{eqn:1-1}) and (\ref{eqn:1-2}).
Suppose that $\bar x$ is a fixed point of eq. (\ref{eqn:1-1}): 
$F(\bar x)=0$ i.e., $f(\bar x)=g(\bar x)$.
Then, $\bar x$ is also a fixed point of eq. (\ref{eqn:1-2}), 
since $F_{\tau}(\bar x)=\bar x\frac{\bar x+\tau f(\bar x)}{\bar x+\tau g(\bar x)}=\bar x$.
On the other hands, if eq. (\ref{eqn:1-2}) has a fixed point $\bar x$, we have $\bar x(f(\bar x)-g(\bar x))=0$.
Thus, the following property is obtained.

\begin{description}
	\item[Prop. 1 (Fixed Point Condition)] \ \\
		A fixed point of eq. (\ref{eqn:1-1}) is identical to 
        a fixed point of eq. (\ref{eqn:1-2}).
%
%
\end{description}
%
%

\subsection{Linear Stability}
\label{sec.2.1}

%
%
%
The stability of the fixed point $\bar x (> 0)$ for eq. (\ref{eqn:1-1}) is determined by 
the following linearized equation,
$
	\frac{dx}{dt} = \frac{dF(\bar{x})}{dx} \cdot x = D(\bar{x}) x, 
$
where 
\begin{eqnarray}
	D(\bar x)=\frac{df(\bar x)}{dx}-\frac{dg(\bar x)}{dx}.
	\label{eqn:D_xbar}
\end{eqnarray}
The stability of $\bar x$ for eq. (\ref{eqn:1-2}) is determined 
by the absolute value of differential coefficient of $F_\tau$ at $\bar x$;
when $|\frac{dF_\tau (\bar x)}{dx}|<1$, $\bar x$ is (asymptotically) stable.  
%
%
%
%
%
%
%
%
At the fixed point $\bar{x}$, the relation $f(\bar x)=g(\bar x)$ holds, and 
the first derivative of $F_\tau$ at $\bar x$ can be represented as
	\begin{eqnarray}
		\frac{dF_\tau (\bar x)}{dx}=1+Z_{\tau}(\bar x)D(\bar x),
		\label{eqn:2-1-1}
	\end{eqnarray}
where 
	\begin{eqnarray}
		Z_\tau (\bar x)=\frac{\tau \bar x}{\bar x +\tau f(\bar x)}.
		\label{eqn:2-1-1a}
	\end{eqnarray}
Then $\bar x$ is stable when 
%
%
%
	\begin{eqnarray}
		-2<Z_\tau(\bar x)D(\bar x)<0.
		\label{eqn:2-1-2}
	\end{eqnarray}
From eq. (\ref{eqn:2-1-1a}), $Z_\tau(\bar x)>0$ always holds since $\bar{x}$, $f$, and $\tau >0$.
On the other hand, the sign of $D(\bar x)$ depends on the value of $\bar x$.
%
%
For $D(\bar x)>0$
we have $Z_\tau(\bar x)D(\bar x)> 0$. 
%
Therefore, when $\bar x$ is unstable in eq. (\ref{eqn:1-1}), 
it is also unstable in eq. (\ref{eqn:1-2}).
For $D(\bar x)<0$, 
$\bar x$ is stable in eq. (\ref{eqn:1-1}).
However, its stability for eq. (\ref{eqn:1-2}) depends 
on $\tau$ and $\kappa (\bar x)$, which is given by   
\begin{eqnarray}
	\kappa(\bar x) = -\frac{2 \bar{x}}{\bar xD(\bar x)+2f(\bar x)}.
	\label{eqn:2-1-3}
\end{eqnarray}
Note that $\kappa (\bar x)$ is independent of $\tau$.
Then, the relation $-2<Z_\tau (\bar x)D(\bar x)<0$ holds for any $\tau>0$ 
when $\kappa (\bar x) < 0$.
%
%
%
%
When $\kappa(\bar x) > 0$, $-2<Z_\tau(\bar x)D(\bar x)$ holds 
only for $\tau $ satisfying $0<\tau<\kappa(\bar x)$.
For $\kappa(\bar x) < \tau$, 
$Z_\tau(\bar x)D(\bar x)<-2$ 
and $\bar x$ is no longer a stable fixed point of eq. (\ref{eqn:1-2}).
Then, relation of linear stability between eqs. (\ref{eqn:1-1}) and (\ref{eqn:1-2}) 
can be summarized as follows.
\begin{description}
	\item[Prop. 2 (Stability Conditions)]
	\begin{description}
		\item[\;]
		
		\item[(a)] When $\bar x$ is a stable fixed point of eq. (\ref{eqn:1-1}), \\
		(a-i) if $\kappa(\bar x) < 0$, $\bar x$ is stable in eq. (\ref{eqn:1-2}) for any $\tau$,\\
 		(a-ii) if $\kappa(\bar x) > 0$ and $0<\tau<\kappa(\bar x)$, 
 		$\bar x$ is stable in eq. (\ref{eqn:1-2}),\\
 		(a-iii) if $\kappa(\bar x) > 0$ and $\tau>\kappa(\bar x)$, 
		$\bar x$ is unstable in eq. (\ref{eqn:1-2}).
		\item[(b)] When $\bar x$ is an unstable 
		fixed point of eq. (\ref{eqn:1-1}),	
			$\bar x$ is also unstable 
			in eq. (\ref{eqn:1-2}) for any $\tau$.
	\end{description}
\end{description}
%
%
%

%
%
Prop. 2 shows that the stable fixed point $\bar x$ for eq. (\ref{eqn:1-1}) 
retains its stability in eq. (\ref{eqn:1-2}) 
when $\tau < \kappa(\bar x)$. 
%
%
When $\bar x$ is unstable 
in eq. (\ref{eqn:1-1}), on the other hand, 
its stability does not change in eq. (\ref{eqn:1-2}) for any $\tau$.

In the case of $\tau \rightarrow \infty$, 
eq. (\ref{eqn:1-2}) becomes 
	\begin{eqnarray}
		x _{n+1}=x _n \frac{f(x _n)}{g(x _n)} \equiv F_\infty (x_n),
		\label{eqn:3-0-1}
	\end{eqnarray}
%
%
%
and the first derivative of $F_\infty$ with respect to $x$ 
at a fixed point $\bar x$ is obtained as 
	\begin{eqnarray}
		\frac{dF_\infty (\bar x)}{dx}=1+Z_{\infty}(\bar x)D(\bar x),
		\label{eqn:3-0-2}
	\end{eqnarray}
where 
	\begin{eqnarray}
		Z_\infty (\bar x)=\frac{\bar x}{f(\bar x)}.
		\label{eqn:3-0-2a}
	\end{eqnarray}
Note that in this limiting case, the condition (a-ii) in Prop. 2 is never satisfied.
Therefore, the stability of $\bar x$ for eq. (\ref{eqn:3-0-1}) can be replaced by the following Prop. 2'.
\begin{description}
	\item[Prop. 2' (Stability Conditions for eq. (\ref{eqn:3-0-1}), 
	the limiting case of $\tau \rightarrow \infty$)] 
	\begin{description}
		\item[\;]

    	\item[(a)] When $\bar x$ is a stable fixed point of eq. (\ref{eqn:1-1}),\\
		(a-i) if $\kappa(\bar x) < 0$, $\bar x$ is stable in eq. (\ref{eqn:3-0-1}), \\
 		(a-ii) if $\kappa(\bar x) > 0$ $\bar x$ is unstable in eq. (\ref{eqn:3-0-1}).
		\item[(b)] When $\bar x$ is an unstable 
		fixed point of eq. (\ref{eqn:1-1}),	
			$\bar x$ is also unstable 
			in eq. (\ref{eqn:3-0-1}).
	\end{description}
\end{description}

\subsection{Flip Bifurcation}
\label{sec.2.2}

The results of (a-ii) and (a-iii) in Prop. 2 
suggest existence of additional bifurcation in eq. (\ref{eqn:1-2}) 
for $\tau$ as a bifurcation parameter.
Let us suppose that $\bar x$ is a positive fixed point of eq. (\ref{eqn:1-2}) with $D(\bar x)<0$ and $\kappa (\bar x) > 0$.
In this case, $\bar x$ becomes nonhyperbolic at $\tau = \kappa (\bar x)$; 
\begin{eqnarray}
	\left. \frac{\partial F_{\tau} (\bar x)}{\partial x} \right|_{\tau=\kappa (\bar x)}=-1 
	\label{eqn:2-2-1}
\end{eqnarray}
holds.
Furthermore, the flip bifurcation can occur 
when eq. (\ref{eqn:1-2}) satisfies the following four conditions\cite{Wiggins} 
at the bifurcation point $(x,\tau)=(\bar x, \kappa (\bar x))$:
\[
\frac{\partial F^2_{\tau} (x)}{\partial \tau}=0, 
\;\:
\frac{\partial^2 F^2_{\tau} (x)}{\partial \tau \partial x}\not =0,
\;\;
\frac{\partial^2 F^2_{\tau} (x)}{\partial x^2}=0,
\;\;
\mbox{and}
\;\;
\frac{\partial^3 F^2_{\tau} (x)}{\partial x^3}\not =0,
\]
where $F^2_{\tau} = F_{\tau} \circ F_{\tau}$.
Among these conditions, the first three conditions
are found to be always satisfied at $(x,\tau)=(\bar x, \kappa(\bar x))$.
Thus, the following proposition is obtained.
\begin{description}
	\item[Prop. 3 (Flip Bifurcation Condition)]
		\ \\
		For a positive fixed point $\bar x$ of eq. (\ref{eqn:1-2}) with $D(\bar x)<0$ and $\kappa (\bar x) > 0$,
		eq. (\ref{eqn:1-2}) exhibits the flip bifurcation 
		at the bifurcation point $\tau = \kappa(\bar x)$
		when $\displaystyle \left.  \frac{\partial^3 F^2_{\tau} (\bar x)}{\partial x^3} \right|_{\tau = \kappa (\bar x)} \not =0$.\\
%
\end{description}
%
%
%
Based on discussion in this subsection, it is also concluded that the additional bifurcation is limited to the flip bifurcation only.

\subsection{Preservation of Original Bifurcations}
\label{sec.2.2a}

Next we consider the case where the original differential equation (\ref{eqn:1-1}) has one of saddle node, transcritical, and supercritical pitchfork bifurcations.
%
%
%
%
We set its bifurcation parameter $c>0$ in eq. (\ref{eqn:1-1})
, which is rewritten as
\begin{equation}
    \frac{dx}{dt} =F(x,c)= f(x,c)-g(x,c), 
    \label{eqn:1-1c}
\end{equation}
where the bifurcation occurs at $(\bar x, \bar c)$.
Applying the tropical discretization to eq. (\ref{eqn:1-1c}),  
we obtain the discrete dynamical system 
\begin{equation}
    x _{n+1} = F_\tau (x_n, c) = x _n \frac{x _n+\tau f(x _n, c)}{x _n+\tau g(x _n, c)}.
    \label{eqn:1-2c}
\end{equation}
Focusing on the following four relations between $F(x,c)$ and $F_\tau (x, c)$ 
at the bifurcation point $(x,c)=(\bar x, \bar c)$, 
\[
\displaystyle\frac{\partial F_{\tau}(\bar x, \bar c)}{\partial c} = Z_{\tau}(\bar x)\frac{\partial F(\bar x, \bar c)}{\partial c}, 
\;\;\; 
\displaystyle\frac{\partial F^2_{\tau}(\bar x, \bar c)}{\partial x^2} = Z_{\tau}(\bar x)\frac{\partial F^2(\bar x, \bar c)}{\partial x^2}, 
\]
\[
\displaystyle\frac{\partial^2 F_{\tau}(\bar x, \bar c)}{\partial x\partial c} = Z_{\tau}(\bar x)\frac{\partial^2 F(\bar x, \bar c)}{\partial x\partial c},
\;\;\;
\displaystyle\frac{\partial^3 F_\tau(\bar x, \bar c)}{\partial x^3} = Z_{\tau}(\bar x)\frac{\partial^3 F(\bar x, \bar c)}{\partial x^3},
\]
we obtain the following proposition for preservation 
of the saddle-node, transcritical, and pitchfork bifurcations 
in eq. (\ref{eqn:1-2c}) based on the bifurcation conditions in eq. (\ref{eqn:1-1c}) \cite{Wiggins}.
\begin{description}
	\item[Prop. 4 (Saddle-node, Transcritical, and Pitchfork Bifurcation Conditions)]
	\begin{description}
		\item[\ ]
		
		\item[(a: saddle-node)] 
		When eq. (\ref{eqn:1-1c}) satisfies the condition 
		for the saddle-node bifurcation at the bifurcation point $(x,c)=(\bar x, \bar c)$: 
		$\frac{\partial F(\bar x, \bar c)}{\partial c}\not=0$ and $\frac{\partial^2 F(\bar x, \bar c)}{\partial x^2}\not =0$, 
		eq. (\ref{eqn:1-2c}) also satisfies the saddle-node bifurcation conditions at $(\bar x,\bar c)$:
		$\frac{\partial F_\tau(\bar x, \bar c)}{\partial c}\not=0$ and $\frac{\partial^2 F_\tau(\bar x, \bar c)}{\partial x^2}\not =0$.
		\item[(b: transcritical)] 
        When eq. (\ref{eqn:1-1c}) satisfies the transcritical bifurcation conditions at $(\bar x, \bar c)$: 
		$\frac{\partial F(\bar x, \bar c)}{\partial c}=0$, 
		$\frac{\partial^2 F(\bar x, \bar c)}{\partial x^2}\not =0$, and $\frac{\partial^2 F(\bar x, \bar c)}{\partial x\partial c}\not =0$,
		eq. (\ref{eqn:1-2c}) also satisfies the transcritical bifurcation conditions at $(\bar x,\bar c)$:
		$\frac{\partial F_\tau(\bar x, \bar c)}{\partial c}=0$, 
		$\frac{\partial^2 F_\tau(\bar x, \bar c)}{\partial x^2}\not =0$, 
		and $\frac{\partial^2 F_\tau(\bar x, \bar c)}{\partial x\partial c}\not =0$.
		\item[(c: supercritical pitchfork)] 
        When eq. (\ref{eqn:1-1c}) satisfies the pitchfork bifurcation conditions at $(\bar x, \bar c)$: 
		$\frac{\partial F(\bar x, \bar c)}{\partial c}=0$, 
		$\frac{\partial^2 F(\bar x, \bar c)}{\partial x^2} =0$, 
		$\frac{\partial^2 F(\bar x, \bar c)}{\partial x\partial c}\not =0$,
		and $\frac{\partial^3 F(\bar x, \bar c)}{\partial x^3} \not=0$,
		eq. (\ref{eqn:1-2c}) also satisfies the pitchfork bifurcation conditions at $(\bar x,\bar c)$:
		$\frac{\partial F_\tau(\bar x, \bar c)}{\partial c}=0$, 
		$\frac{\partial^2 F_\tau(\bar x, \bar c)}{\partial x^2} =0$, 
		$\frac{\partial^2 F_\tau(\bar x, \bar c)}{\partial x\partial c}\not =0$,
		and $\frac{\partial^3 F_\tau(\bar x, \bar c)}{\partial x^3} \not=0$.		
    \end{description}
\end{description}
%
%
%
%
%
Note that Props. 3 and 4 reproduce  
the dynamical relations in the non-standard finite difference schemes\cite{Alexander2005}. 
It is also noted that 
Prop. 4 holds even in the case $\tau \to \infty$.
%
Then the saddle-node, transcritical, and pitchfork bifurcations in the original continuous differential equations are retained in the limit of $\tau \to \infty$ for the tropically discretized equations.
%

For occurrence condition of the flip bifurcation, 
eq. (\ref{eqn:1-2c}) also satisfies Prop. 3.
Therefore, when $\displaystyle D(\bar x, c) = \frac{\partial f(\bar x, c)}{\partial x}-\frac{\partial g(\bar x, c)}{\partial x}<0$ and $\displaystyle \kappa (\bar x, c) = -\frac{2 \bar{x}}{\bar xD(\bar x, c)+2f(\bar x, c)} > 0$,
eq. (\ref{eqn:1-2c}) exhibits the flip bifurcation 
at $\tau = \kappa(\bar x, c)$
when $\displaystyle\frac{\partial^3 F^2_{\tau} (\bar x, c)}{\partial x^3} \not =0$.
Especially in the case of $\tau \to \infty$, 
it is found that occurrence of the flip bifurcation depends on the sign of $\kappa$, namely, the sign of $\bar x D(\bar x, c) + 2f(\bar x, c)$.
Then there exists another bifurcation point $c^{\ast}$ for the flip bifurcation, where $c^{\ast}$ satisfies $2f(\bar x, c^{\ast})+\bar x D(\bar x, c^{\ast})=0$.

%
%
%
%
%
%
%


\section{Examples}
\label{sec.3}

%
%
%
%
%


\subsection{Flip bifurcation}
\label{sec.3.1}

%
Let us consider the following simple differential equation for $x(t)>0$:  
	\begin{eqnarray}
  		\frac{dx}{dt} = -4x^3+x^2-x+4.
  		\label{eqn:3-1-1}
	\end{eqnarray}
Equation (\ref{eqn:3-1-1}) has a unique stable fixed point $\bar x=1$; 
any $x$ converges to this point monotonically as $t \rightarrow +\infty$.
Dividing the right hand side of eq. (\ref{eqn:3-1-1}) 
into the positive part $f(x)=x^2+4$ and the negative part $g(x)=4x^3+x$, 
the tropically discretized equation for eq. (\ref{eqn:3-1-1}) becomes   
	\begin{equation}
		x_{n+1}=F_\tau (x_n) \equiv x_n \frac{x_n+\tau f(x_n)}{x_n+\tau g(x_n)} = \frac{x_n+\tau (x_n^2+4)}{1+\tau (4x_n^2+1)}.
	\label{eqn:3-1-2}
	\end{equation}
From Prop. 1, $\bar x=1$ is found to be a fixed point of eq. (\ref{eqn:3-1-2}), and we obtain $D(\bar x)= \frac{df(\bar x)}{dx}-\frac{dg(\bar x)}{dx}=-11<0$ 
and $\kappa (\bar x)=2 > 0$.
Then from Prop. 2, the stability of $\bar x$ depends on the value of $\tau$; $\bar x$ is stable (unstable) when $\tau <2$ ($\tau >2$). 
Furthermore, since $\frac{\partial F_{\bar \tau} (\bar x)}{\partial x}=-1$ and 
$	\frac{\partial^3 F^2_{\bar \tau} (\bar x)}{\partial x^3}=-\frac{432}{121}\not =0$,
Prop. 3 tells us that eq. (\ref{eqn:3-1-2}) exhibits a flip bifurcation at $(\bar x, \bar \tau)=(1,2)$.
Note that $\frac{\partial F_{\tau} (\bar x)}{\partial x}=0$ if and only if $\tau =1/6$.

\begin{figure}[h!]
	\begin{center}
	\includegraphics[width=5cm]{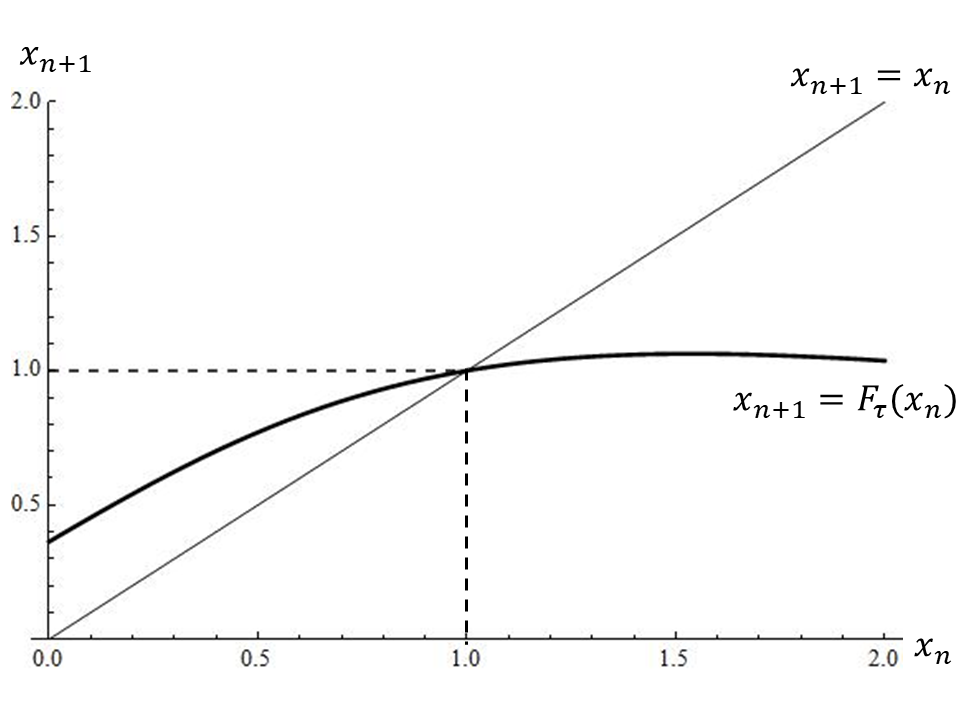}
	\hspace{1mm}
	\includegraphics[width=5cm]{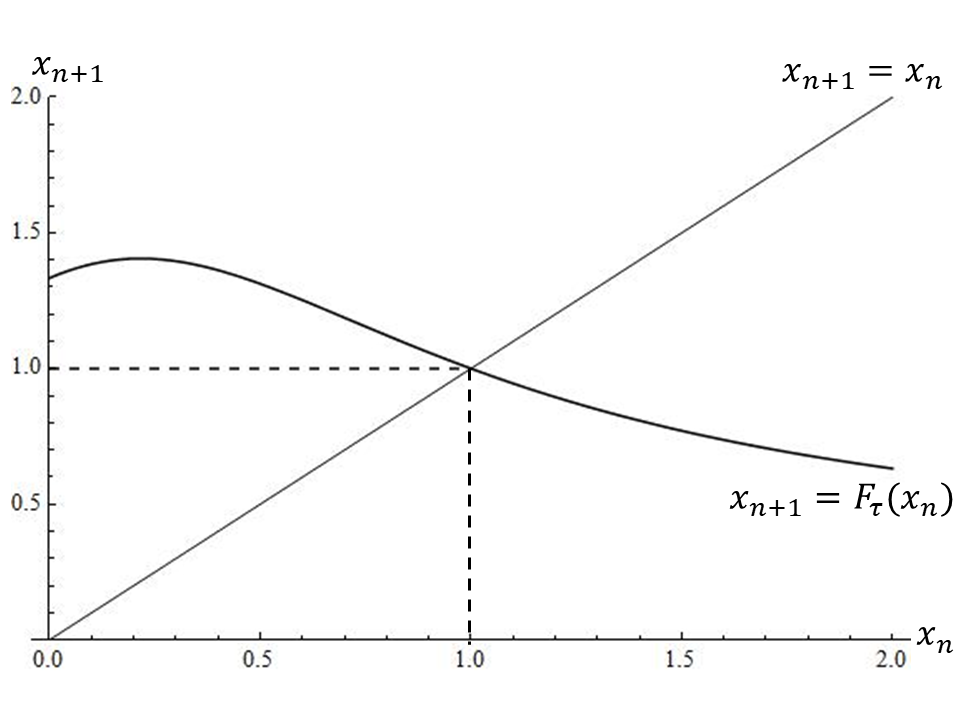}
	\hspace{1mm}
	\includegraphics[width=5cm]{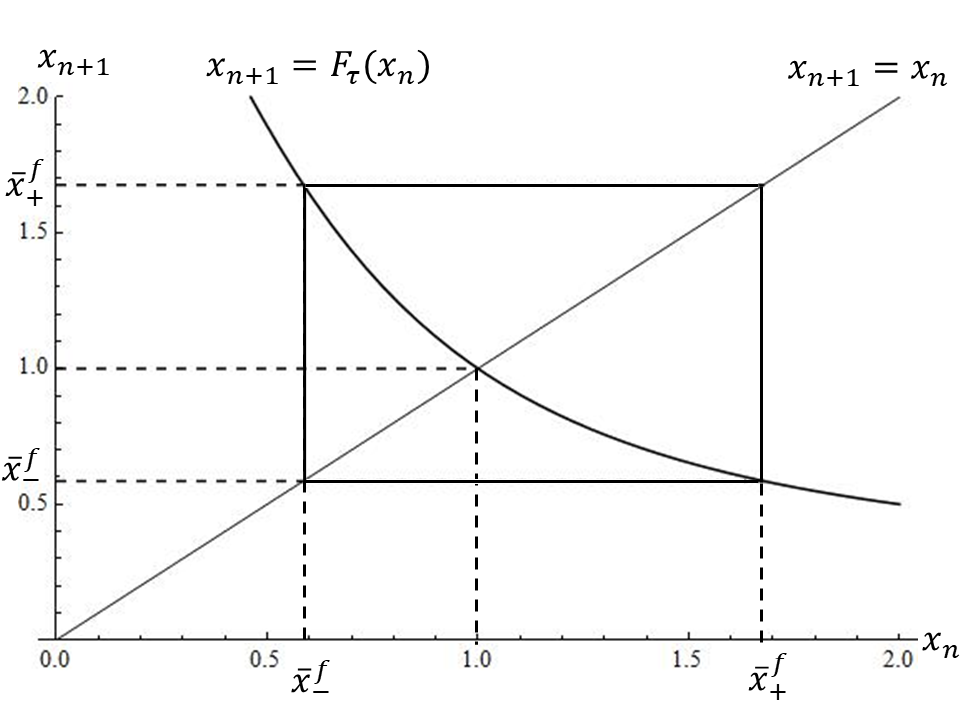}
	\\
	(a)
	\hspace{4cm}
	(b)
	\hspace{4cm}
	(c)
	\caption{\label{Fig.flip_graph} The graphs of eq. (\ref{eqn:3-1-2}). (a) $\tau =0.1$, (b) $\tau =0.5$, and (c) $\tau =3$.}
	\end{center}
\end{figure}

\begin{figure}[h!]
	\begin{center}
	\includegraphics[width=7cm]{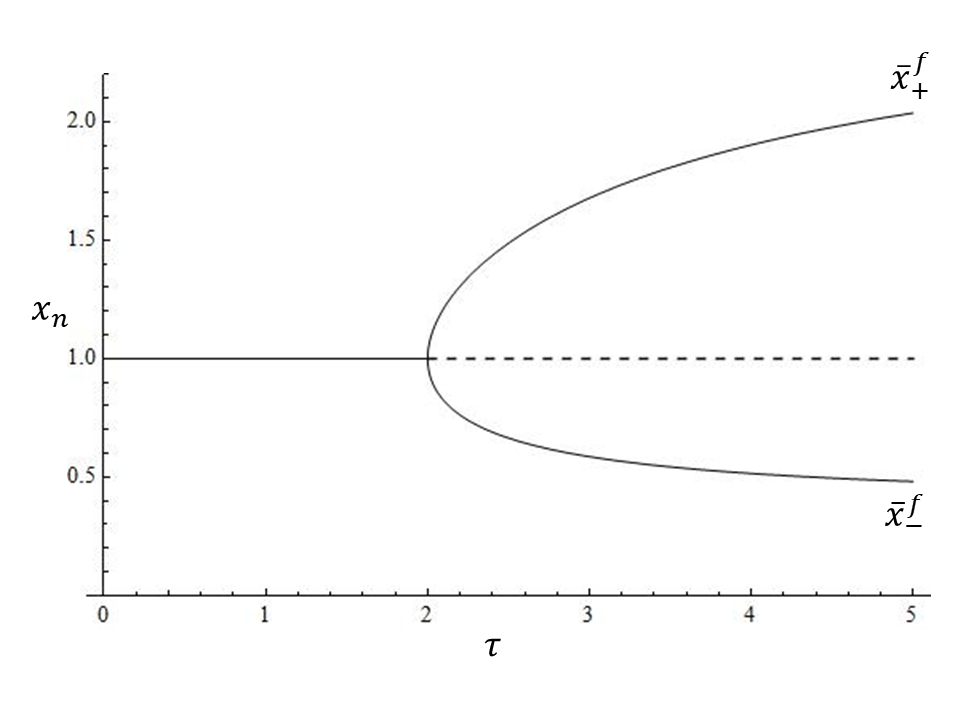}
	\caption{\label{Fig.flip_diagram} The bifurcation diagram for the flip bifurcation of eq. (\ref{eqn:3-1-2}). }
	\end{center}
\end{figure}

Figure \ref{Fig.flip_graph} shows the graphs of eq. (\ref{eqn:3-1-2}) 
for (a) $\tau=0.1$, (b) $\tau=0.5$, and (c) $\tau =3$. 
For the case (a), 
the graph of eq. (\ref{eqn:3-1-2}) intersects the diagonal $x_{n+1} = x_{n}$ at the stable fixed point $\bar{x}=1$.
Then, any initial state converges monotonically to this point 
and eq. (\ref{eqn:3-1-2}) retains the dynamics of eq. (\ref{eqn:3-1-1}).
For the case (b), 
the fixed point $\bar{x}=1$ is still stable, although different from the case (a), 
it becomes a stable focus 
due to $-1<\frac{\partial F_{\tau} (\bar x)}{\partial x}<0$.
For the case (c), 
the fixed point $\bar{x}$ becomes unstable 
and an attracting cycle $\mathcal{C}_f=\{ \bar x^f_{+},\bar x^f_{-}\}$ with period $2$ emerges, where 
$\bar x^f_{\pm}=\frac{1}{2(4\tau+5\tau^2)}
(-2\tau +15\tau^2 \pm \sqrt{-32\tau-84\tau^2-200\tau^3+125\tau^4})$.
The cycle $\mathcal{C}_f$ surrounds the unstable fixed point and 
any $x_n$ starting from $x_0\not =1$ finally arrives at $\mathcal{C}_f$.
Figure \ref{Fig.flip_diagram} shows the bifurcation diagram; the flip bifurcation occurs at $\tau=2$.
%
%


\subsection{Saddle-node bifurcation}
\label{sec.3.2}

%
For the saddle-node bifurcation, 
the following nonlinear differential equation 
for positive $x(t)$ is considered:
\begin{eqnarray}
  \frac{dx}{dt} = c+x(x-2), 
  \label{eqn:3-2-1}
\end{eqnarray}
where $c$ is the bifurcation parameter.
In eq. (\ref{eqn:3-2-1}), the saddle-node bifurcation occurs 
at $(\bar x, \bar c)=(1,1)$.
When $0<c<1$, there are two fixed points 
$\bar x_{\pm} = 1 \pm \sqrt{1-c}$, 
where $\bar x_{-}$ and $\bar x_{+}$ are stable and unstable, respectively.
When $c>1$, eq. (\ref{eqn:3-2-1}) has no fixed points.

The tropically discretized equation for eq. (\ref{eqn:3-2-1}) is 
	\begin{eqnarray}
		x_{n+1}=F_\tau (x_n,c) = \frac{x_n+\tau (c+x_n^2)}{1+2\tau }.
	\label{eqn:3-2-2}
	\end{eqnarray}
From Prop. 4 (a), eq. (\ref{eqn:3-2-2}) also exhibits the saddle-node bifurcation at $(\bar x, \bar c)=(1,1)$.
%
%
Furthermore, from Props. 1 and 2 (b), 
$\bar x_{+}$ also becomes the unstable fixed points 
of eq. (\ref{eqn:3-2-2}) for any $\tau$.
For $\bar x_{-}$, 
\[\kappa(\bar{x}_{-}) = -\frac{2\bar{x}_{-}}{\bar{x}_{-}D(\bar{x}_{-})+2f(\bar{x}_{-})}
= -\frac{\bar x _{-}}{2\bar x _{-}^2-\bar x _{-}+c}
= -\frac{1}{2-\sqrt{1-c}}
\]
becomes negative for $c<1$, then 
$\bar x_{-}$ is found to be the stable fixed points of eq. (\ref{eqn:3-2-2}) for any $\tau$ from Prop. 2 (a-i).
%
%
Therefore, eq. (\ref{eqn:3-2-2}) has only the saddle-node bifurcation,
whose bifurcation diagram coincides with of the original differential equation (\ref{eqn:3-2-1}).
%
%
%
%
%
%
%
%
%
%
%
\subsection{Transcritical bifurcation}
\label{sec.3.3}

For the transcritical bifurcation, we begin with  
\begin{eqnarray}
  \frac{dx}{dt} = (x-1)(c-x),
  \label{eqn:3-3-1}
\end{eqnarray}
where we consider the case $x(t)>0$ and $c>0$.
For this equation, the transcritical bifurcation occurs 
at $(\bar x, \bar c)=(1,1)$.
When $0<c<1$, eq. (\ref{eqn:3-3-1}) has the stable and unstable fixed points $\bar x=1$ and $\bar x=c$, respectively.
And when $c > 1$, $\bar x=1$ and $\bar x=c$ become unstable and stable. 

By tropical discretization of eq. (\ref{eqn:3-3-1}), we obtain 
	\begin{eqnarray}
		x_{n+1}= F_\tau (x_n,c)= x_n\frac{x_n+\tau (1+c)x_n}{x_n+\tau (x_n^2+c)}.
	\label{eqn:3-3-2}
	\end{eqnarray}
From Prop. 4 (b), it is found that eq. (\ref{eqn:3-3-2}) 
also exhibits the transcritical bifurcation.
When $0<c<1$, it is found from Prop. 2 (b) that $\bar x=c$ becomes the unstable fixed point of eq. (\ref{eqn:3-3-2}) for any $\tau$.
And from $\kappa (\bar x=1)=-\frac{2}{1+3c} < 0$,  $\bar x=1$ is the stable fixed point for any $\tau$.
Similarly when $c > 1$, 
it is confirmed that $\bar x=c$ is stable and $\bar x=1$ is unstable 
for any $\tau$.
Therefore, eq. (\ref{eqn:3-3-2}) has only the transcritical bifurcation, and its bifurcation diagram is the same as that of eq. (\ref{eqn:3-3-1}).
%

\subsection{Pitchfork bifurcation}
\label{sec.3.4}

For the supercritical pitchfork bifurcation, 
we consider 
\begin{eqnarray}
  \frac{dx}{dt} = 3cx(x-1)+1-x^3.
  \label{eqn:3-4-1}
\end{eqnarray}
Here, $c$ is the positive bifurcation parameter and the supercritical pitchfork bifurcation occurs at $(\bar x, \bar c)=(1,1)$.
When $c>1$, eq. (\ref{eqn:3-4-1}) has three fixed points $\bar x=1, \bar x_{\pm} = \frac{3c-1 \pm \sqrt{ (1-3c)^2-4}}{2}$; 
$\bar x=1$ is unstable and $\bar x_{\pm}$ are stable.
When, $0<c<1$, this equation has one unique stable fixed point 
$\bar x=1$.

By the tropical discretization of eq. (\ref{eqn:3-4-1}), 
we obtain  
	\begin{eqnarray}
		x_{n+1}= F_\tau (x_n,c) =\frac{x_n+\tau (3cx_n^2+1)}{1+\tau (x_n^2+3c)}.
	\label{eqn:3-4-2}
	\end{eqnarray}
It is found from Prop. 4 (c) that the supercritical pitchfork bifurcation occurs at $(\bar x, \bar c)=(1,1)$.
%
%
%
%
Furthermore, as mentioned below, eq. (\ref{eqn:3-4-2}) possesses the flip bifurcation for $\tau$.
When $c>1$, from Prop. 1, 
eq. (\ref{eqn:3-4-2}) has three fixed points $\bar x=1, \bar x_{\pm}$.
As for their stabilities, from Prop. 2 and $\kappa (\bar x_{\pm}) < 0$, 
$\bar x=1$ is unstable and $\bar x_{\pm}$ are stable for any $\tau > 0$.
%
%
When $0<c<1$, from Prop. 1, 
$\bar x=1$ is also the only fixed point for eq. (\ref{eqn:3-4-2}).
However, the stability of $\bar x$ for $0<c<1$ varies 
with the sign of $\kappa (\bar x) = -\frac{2}{9c-1}$.
%
Actually when $\frac{1}{9}\leq c$, $\kappa (\bar x) < 0$.
Then from Prop. 2, $\bar x$ is stable for any $\tau >0$.
%
%
If $0<c<\frac{1}{9}$, on the other hand, $\kappa (\bar x)$ becomes positive.
Then, the fixed point $\bar x=1$ becomes stable and unstable for $\tau < \kappa (\bar x)$ and $\tau > \kappa (\bar x)$, respectively.

\begin{figure}[h!]
	\begin{center}
	\includegraphics[width=9cm]{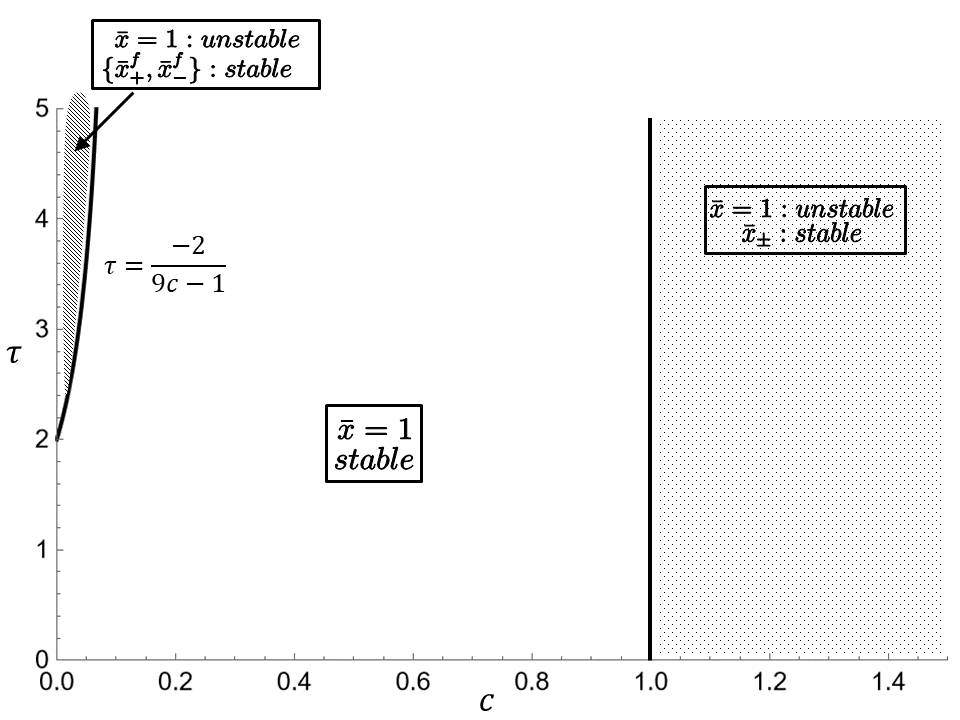}\\
	\caption{\label{Fig.pitchfork_parameters} $\tau$-$c$ diagram for eq. (\ref{eqn:3-4-2}).
	$c=1$ and $\tau=\frac{-2}{9c-1}$ are the supercritical pitchfork and the flip bifurcation curves, respectively. }
	\end{center}
\end{figure}

\begin{figure}[h!]
	\begin{center}
	\includegraphics[width=6cm]{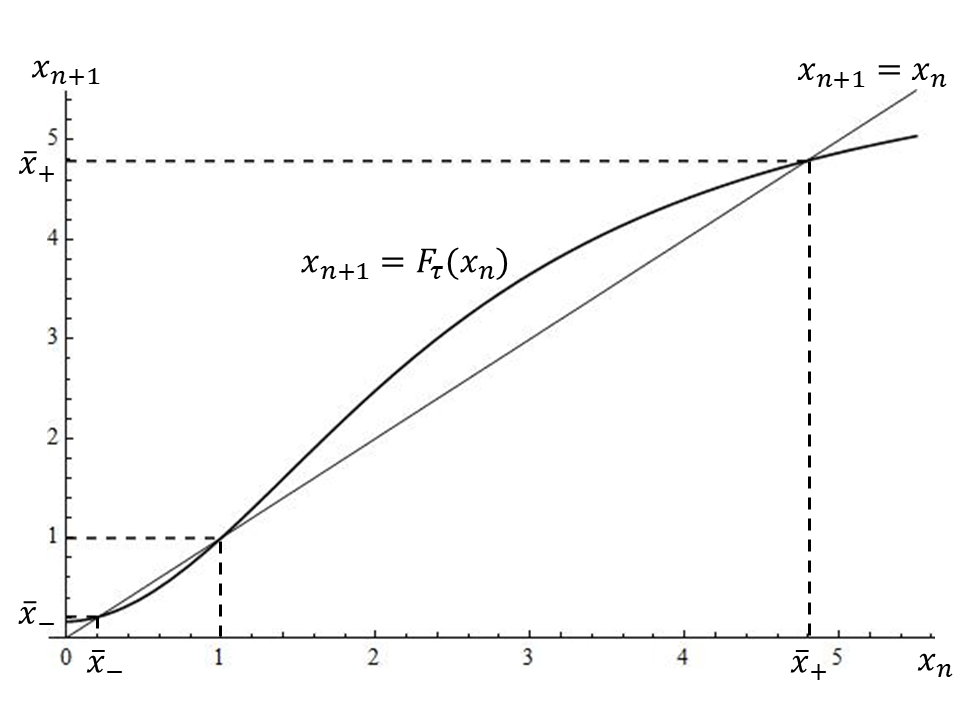}
	\hspace{2mm}
	\includegraphics[width=6cm]{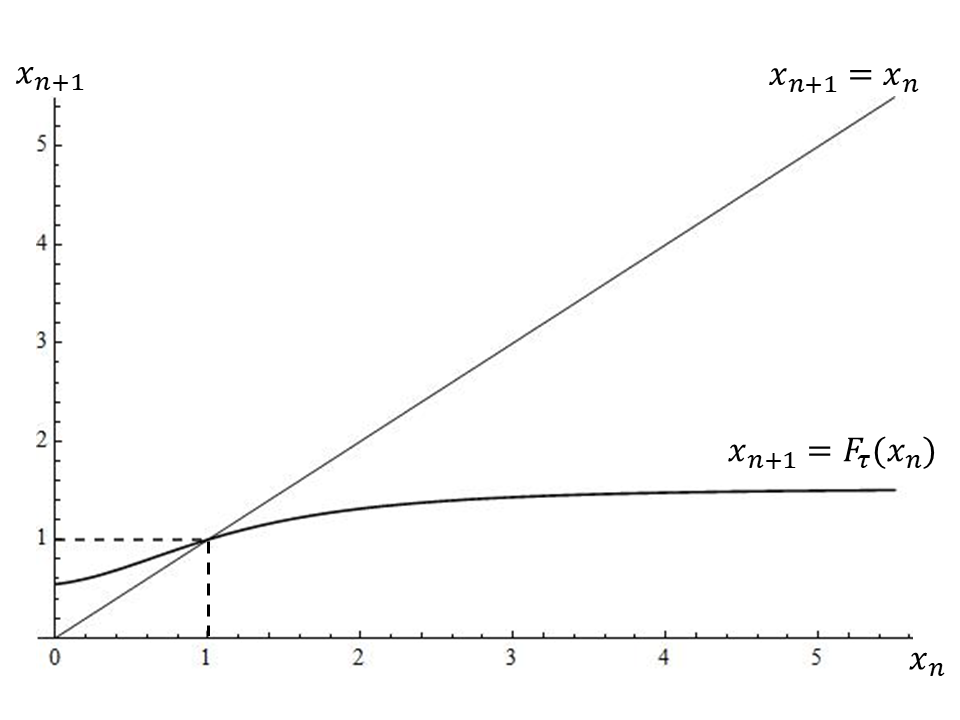}
	\\
	(a)
	\hspace{6cm}
	(b)
	\\
	\includegraphics[width=6cm]{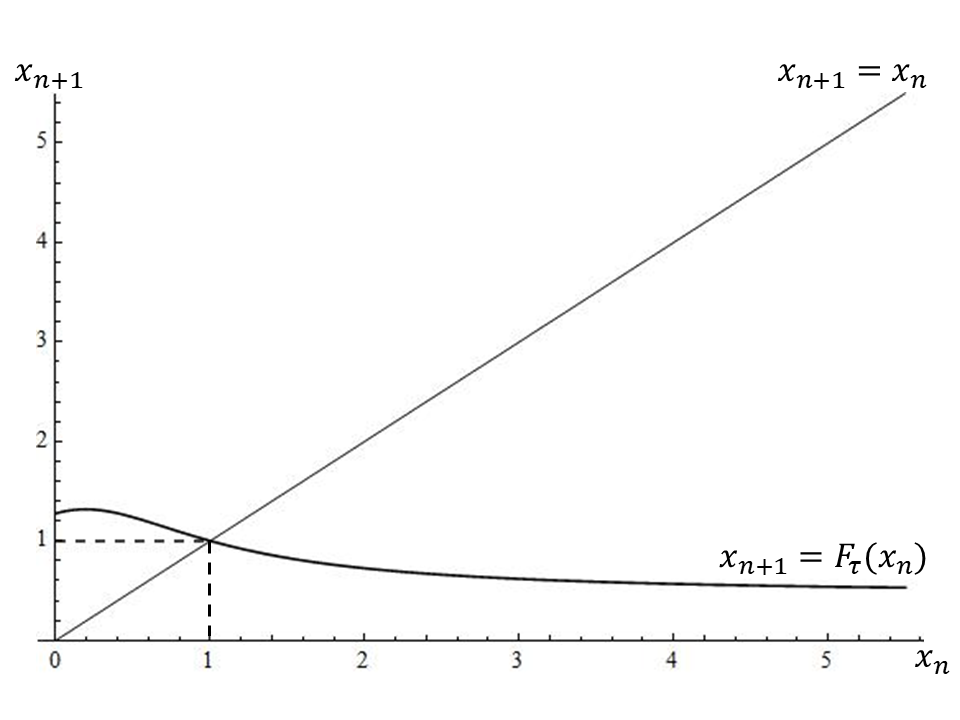}
	\hspace{1cm}
	\includegraphics[width=6cm]{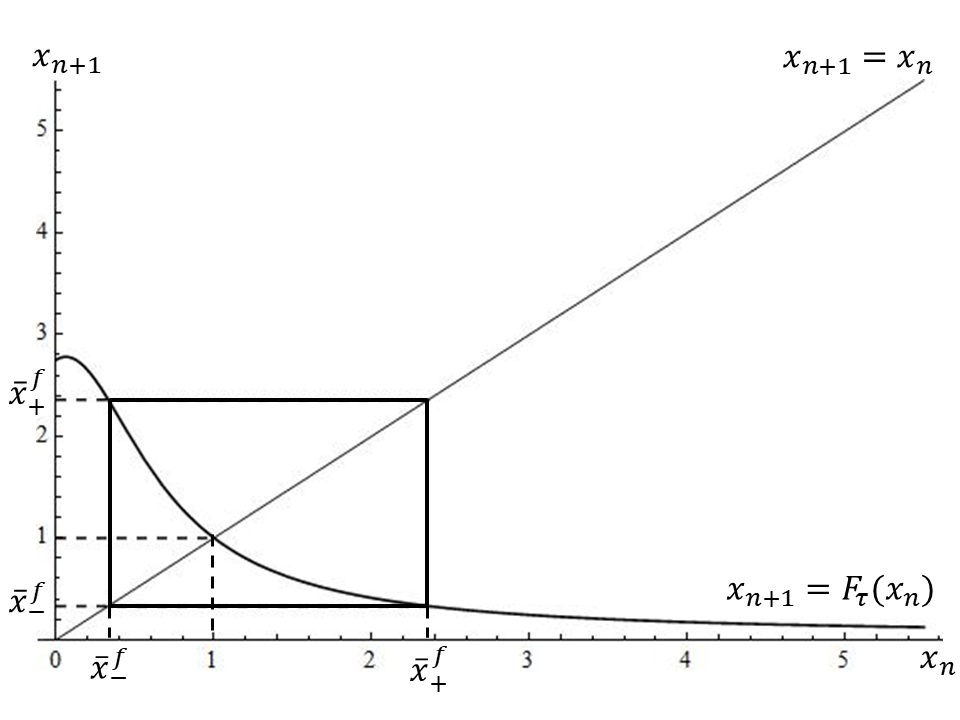}
	\\
	(c)
	\hspace{6cm}
	(d)
	\\
	\caption{\label{Fig.pitchfork_graph} 
		The graphs of eq. (\ref{eqn:3-4-2}) for $\tau=3$. (a) $c =2$, (b) $c =0.5$, (c) $c =0.15$, and (d) $c =0.01$.
					}
	\end{center}
\end{figure}

\begin{figure}[h!]
	\begin{center}
	\includegraphics[width=7cm]{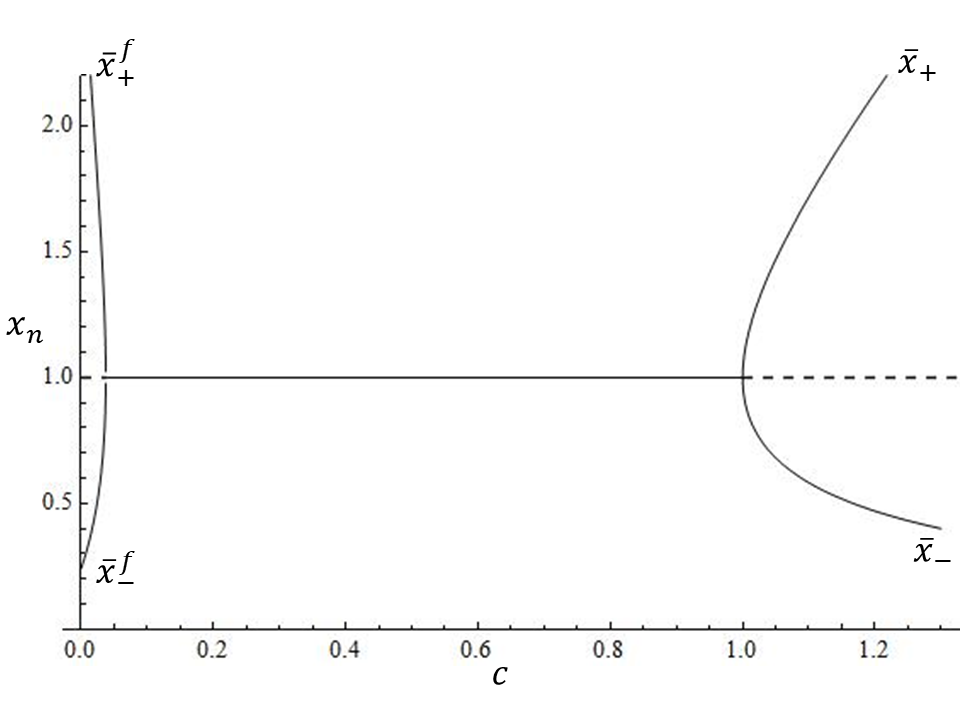}
	\caption{\label{Fig.pitchfork_diagram} The bifurcation diagram of eq. (\ref{eqn:3-4-2}) for $\tau =3$. }
	\end{center}
\end{figure}

Considering $\frac{\partial^3 F^2_{\bar \tau} (\bar x, c)}{\partial x^3}= -\frac{8(18c^2-3c+1)}{3(c-1)^2} \ne 0$ for $ 0<c<\frac{1}{9}$, 
it is found from Prop. 3 that the flip bifurcation occurs 
at $\tau=\kappa (\bar x)$.
Actually for $\tau > \kappa (\bar x)$, 
there is an attracting cycle $\mathcal{C}^f=\{\bar x^f_{+}, \bar x^f_{-}\}$ with period 2 
around $\bar x$, where
\[
\bar x^f_{\pm}=\frac{-6c\tau+\tau^2-9c^2\tau^2
\pm\sqrt{(6c\tau-\tau^2+9c^2\tau^2)-4(\tau +3c\tau ^2+9c^2\tau ^2)(2+9c\tau +3c \tau^2 +9c^2 \tau^2)}}{2(\tau+3c\tau^2+9c^2\tau^2)}.
\]
%
%
%
Figure \ref{Fig.pitchfork_parameters} shows $\tau$-$c$ diagram for the dynamics of eq. (\ref{eqn:3-4-2}).
$c=1$ and $\tau=\frac{-2}{9c-1}$ are the bifurcation curves 
on which the supercritical pitchfork and the flip bifurcations occur, respectively.
%
%
Figure \ref{Fig.pitchfork_graph} shows the graphs of eq. (\ref{eqn:3-4-2}) with $\tau=3$ 
for (a) $c=2$, (b) $c =0.5$, (c) $c =0.15$, and (d) $c=0.01$. 
For (a), 
from intersection of the curve and the diagonal $x_{n+1}=x_{n}$, 
it is found that there are three fixed points $x_n=1,\bar x_{\pm}$, 
where $\bar x=1$ is unstable and $\bar x_{\pm}$ are stable.
For (b) and (c), the curves intersect the diagonal only at the stable fixed point $\bar x$, which becomes node in (b) and focus in (c).
For (d), the fixed point $\bar x$ becomes unstable
and an attracting cycle $\mathcal{C}_f=\{ \bar x^f_{+},\bar x^f_{-}\}$ with period $2$ emerges.
%
%
Figure \ref{Fig.pitchfork_diagram} shows the bifurcation diagram for $\tau=3$; 
the supercritical pitchfork bifurcation occurs at $c=1$ and
the flip bifurcation occurs at $c=\frac{1}{27}$.
%
%
%
%
%
%
%
%
%

Perviously we also treated the following tropically discretized equation instead of eq. (\ref{eqn:3-4-2}) for the supercritical pitchfork bifurcation\cite{Ohmori2020},
	\begin{eqnarray}
		x_{n+1}  = F_{\tau}(x_n) = 
				\begin{cases}
					\displaystyle
                  \frac{x_n+\tau (3cx_n^2+1))}{1+\tau (x_n^2+3c)}
                  \;\;\;\; \mbox{for  } c \geq 1, 
		\\
		\displaystyle
                  \frac{x_n+\tau\{ 3 x_n^2 + \eta + 1 \}}
                  {1+\tau\{x_n^2 + 3\eta x_n + 3\}}
                  \;\;\;\; \mbox{for  } 0<c < 1,
		\label{eqn:3-5-1}
				\end{cases}
	\end{eqnarray}
where $\eta =1-c (>0)$.
%
%
%
In this equation, 
from $\kappa(\bar x)=-\frac{2}{8+3\eta} <0$ for $0<c<1$, 
this fixed point is found to be stable for any $\tau$.
%
Therefore, eq. (\ref{eqn:3-5-1}) exhibits only the supercritical pitchfork bifurcation at $(\bar x, \bar c)=(1,1)$, 
and the flip bifurcation does not occur; the dynamical property of eq. (\ref{eqn:3-5-1}) is consistent 
with that of eq. (\ref{eqn:3-4-1}) at any $\tau$.
This result suggests that by appropriate tropical discretization, it is possible to construct a discrete dynamical system having only the same bifurcation point as the original continuous one.
%

\section{Correspondence with ultradiscrete dynamical systems}
\label{sec:4}

%
Here, we consider correspondence between the tropically discretized dynamical systems and their ultradiscrete ones 
by focusing the cases of saddle-node and supercritical pitchfork bifurcations.
%
%

\subsection{Saddle-node bifurcation}
\label{sec.4.1}

In the previous paper, we derived the following max-plus equation for saddle-node bifurcation\cite{Ohmori2020}, 
\begin{equation}
    X_{n+1} = \max(PX_n, C),
    \label{eqn:4-1a}
\end{equation}
where $P>1$ and $C$ is a bifurcation parameter; 
its bifurcations occurs at $(\bar X, \bar C)=(0,0)$.

From eq. (\ref{eqn:3-2-2}), on the other hand, we obtain the tropically discretized equation 
%
	\begin{eqnarray}
		x_{n+1}=  \frac{c+x_n^2}{2}
	\label{eqn:4-1-1}
	\end{eqnarray}
when $\tau \to \infty$.
From Props. 2' and 4 (a) in Sec. 2, 
it is found that eq. (\ref{eqn:4-1-1}) shows the saddle-node bifurcation at $(\bar x,\bar c)=(1,1)$.
Adopting the variable transformations
\begin{eqnarray}
	x_n=e^{X_n/\varepsilon} 
	\mbox{ and  } c=e^{C/\varepsilon}, 
\label{eqn:40}
\end{eqnarray}
eq. (\ref{eqn:4-1-1}) can be rewritten as 
\begin{eqnarray}
	X_{n+1}= \varepsilon \log(e^{C/\varepsilon}+ e^{2X_n/\varepsilon})-\varepsilon \log 2, 
\label{eqn:4-1-2}
\end{eqnarray}
%
%
which also possesses the saddle-node bifurcation at $(\bar X,\bar C)=(0,0)$.
%
%
%
%
%
Considering the ultradiscrete limit, $\varepsilon \to +0$, of eq. (\ref{eqn:4-1-2}), we obtain from eq. (\ref{eqn:0})   
	\begin{eqnarray}
		X_{n+1}= \max(C, 2X_n), 
	\label{eqn:4-1-3}
	\end{eqnarray}
which is identical to eq. (\ref{eqn:4-1a}) with $P=2$.

It is noted that ultradiscretization brings about piecewise linearization of eq. (\ref{eqn:4-1-2}) by taking the limit of zooming out for the scale transformations with respect to $X$, $C$, and 
the function $S(X, C)$ given as
\begin{equation}
    S(X, C) = \log(e^{C}+ e^{2X})-\log 2.
\end{equation}
By using $S(X, C)$, eq. (\ref{eqn:4-1-2}) can be rewritten as 
$X_{n+1}= \varepsilon S(X_{n}/\varepsilon, C/\varepsilon)$.
Therefore, if there exists $(X^{\ast}, C^{\ast})$ 
satisfying  
$S(X, C) = \varepsilon S(X/\varepsilon, C/\varepsilon)$ 
for any $\varepsilon$, then $(X^{\ast}, C^{\ast})$ 
exhibits scale invariance.
Actually, in this case, we obtain  
$(X^{\ast}, C^{\ast}) = (C/2, C)$.
%
%
Furthermore, since $\displaystyle \frac{\partial S}{\partial X}$ is a monotonically increasing function 
from $0$ to $2$ and $\displaystyle \frac{\partial S}{\partial X}(X^{\ast}, C^{\ast}) = 1$ for any $\varepsilon$, even in the limit of $\varepsilon \to +0$, eq. (\ref{eqn:4-1-3}) retains the same saddle-node bifurcation as eq. (\ref{eqn:4-1-1}) and eq. (\ref{eqn:3-2-1}).

%
%

%
%

%
%
%


%

\subsection{Supercritical pitchfork and flip bifurcations}
\label{sec.4.2}

Next, we consider the following max-plus equation for supercritical pitchfork bifurcation\cite{Ohmori2020}, 
\begin{equation}
    X_{n+1}=\max(PX_n+C,0)-\max(PX_n,C).
    \label{eqn:4-1c}  
\end{equation}
Here $P>1$, and $C$ is a bifurcation parameter.
and the bifurcation occurs at $(\bar X, \bar C)=(0,0)$.

From eq. (\ref{eqn:3-4-2}), we obtain the tropically discretized equation 
%
%
%
\begin{eqnarray}
	x_{n+1}=  \frac{3cx_n^2+1}{x_n^2+3c},
\label{eqn:4-3-1}
\end{eqnarray}
as the limit of $\tau \to \infty$.
After the variable transformations (\ref{eqn:40}), 
we obtain 
	\begin{eqnarray}
		X_{n+1}= \varepsilon \log(3e^{(2X_n+C)/\varepsilon}+ 1)
		-\varepsilon \log (e^{2X_n/\varepsilon}+ 3e^{C/\varepsilon}).
	\label{eqn:4-3-2}
	\end{eqnarray}
From Props. 2', 3, and 4 (c) in Sec. 2, eq. (\ref{eqn:4-3-2}) also possesses the same supercritical pitchfork and flip bifurcations as eq. (\ref{eqn:4-3-1}) at $(\bar X, \bar C)=(0,0)$.
Considering the ultradiscrete limit, $\varepsilon \to +0$, of eq. (\ref{eqn:4-3-2}), we obtain    
	\begin{eqnarray}
		X_{n+1}= \max(2X_n+C,0)-\max(2X_n,C), 
	\label{eqn:4-3-3}
	\end{eqnarray}
which coincides with eq. (\ref{eqn:4-1c}) for $P=2$. 
%
%
Equation (\ref{eqn:4-3-3}) shows that  ultradiscretization brings about piecewise linearization of eq. (\ref{eqn:4-1-2}) as in the case of Sec. \ref{sec.4.1}.

Equation (\ref{eqn:4-3-2}) can be also rewritten as 
$X_{n+1}= \varepsilon S(X_{n}/\varepsilon, C/\varepsilon)$ when we set 
\begin{equation}
    S(X, C) = \log(3e^{2X+C}+ 1)- \log (e^{2X}+ 3e^{C}).
\end{equation}
Then, $(X^{\ast}, C^{\ast})=(0, C)$ in this case, 
where $X^{\ast}=0$ is one of the fixed points of eq. (\ref{eqn:4-3-2}).
It is noted that 
$\displaystyle \frac{\partial S}{\partial X}(X=\pm \infty, C/\varepsilon) = 0$.
When $3e^{C/\varepsilon} > 1$, $\varepsilon S(X_{n}/\varepsilon, C/\varepsilon)$ becomes 
monotonically increasing function of $X$.
When $3e^{C/\varepsilon} < 1$, on the other hand, 
it becomes monotonically decreasing function.
Therefore in the limit of $\varepsilon \to +0$, 
the crossover between 
the supercritical bifurcation ($C>0$) and 
the flip bifurcation ($C<0$) occurs at $C=0$, 
where $\displaystyle \frac{\partial S}{\partial X}(X=0, C/\varepsilon) = -2, 1, +2$ when $C$ is negative, zero, and positive, respectively.

Here we note that dynamical properties of tropically discretized equations depend on how $F(x)$ is divided into the positive smooth functions $f(x)$ and $g(x)$ in the original differential equation (\ref{eqn:1-1}).
Then the propositions shown in Sec. \ref{sec.2} are available for verification of appropriate choice of $f(x)$ and $g(x)$ whether the original bifurcations retain or additional flip bifurcations emerge.
%
%
%
%
So far in this paper, we have focused only on one-dimensional cases.
Our approach is considered to be extended in cases of higher dimensions, e.g., the two dimensional dynamical systems having limit cycles by Hopf bifurcation\cite{Ohmori2021,Isojima2022,Matsuya2019} and the dynamical systems with spatial dependence\cite{Murata2013,Matsuya2015,Ohmori2016}.
%
%

\section{Conclusion}
\label{sec:5}

In this paper, we have shown the general propositions of stability and bifurcation properties for the tropically discretized dynamical systems, which are derived from the one dimensional continuous dynamical system.
We have also identified the occurrence condition of the additional flip bifurcation for the discretized time interval $\tau$ introduced in tropical discretization.
Some application examples have been demonstrated.
Especially even in the case of $\tau \to \infty$, the tropically discretized dynamical systems retains the bifurcations of the original continuous dynamical system.
Furthermore, we can derive the max-plus equations via ultradiscretization from the tropically discretized equations for $\tau \to \infty$; 
ultradiscretization plays a role of piecewise linearization.
We have shown that the derived max-plus equations also retain the bifurcations of the original continuous dynamical system.
Therefore, it is concluded that the tropical discretization and the ultradiscretization can link the continuous differential descriptions and ultradiscrete max-plus ones for one dimensional dynamical systems.

\bigskip

\noindent
{\bf Acknowledgement}

The authors are grateful to 
Prof. D. Takahashi, 
Prof. T. Yamamoto, and Prof. Emeritus A. Kitada 
at Waseda University, 
Associate Prof. K. Matsuya at Musashino University, Prof. M. Murata at Tokyo University of Agriculture and Technology 
for useful comments and encouragements. 
This work was supported by JSPS
KAKENHI Grant Numbers 22K13963 and 22K03442.

\bigskip

\noindent
{\bf Data Availability}

Data sharing is not applicable to this article as no new data were created or analyzed in this study.

\end{document}